\documentclass{llncs}

\usepackage{llncsdoc}
\usepackage{graphicx}

\newcommand{\keywords}[1]{\par\addvspace\baselineskip
\noindent\keywordname\enspace\ignorespaces#1}

\begin {document}
\title{Fast Verifying Proofs of Propositional Unsatisfiability via Window Shifting}
\titlerunning{Fast Verifying Proofs of Unsatisfiability}

\author{Jingchao Chen}
\institute{School of Informatics, Donghua University \\
2999 North Renmin Road, Songjiang District, Shanghai 201620, P. R.
China \email{chen-jc@dhu.edu.cn}}

\maketitle
\begin{abstract}
The robustness and correctness of SAT solvers are receiving more and more attention.
In recent SAT  competitions, a proof of unsatisfiability emitted by SAT solvers must
be checked. So far, no proof checker has been efficient for every case. In the SAT competition 2016,
some proofs were not
verified within 20000 seconds. For this reason, we decided to develop a more efficient proof checker called TreeRat.
This new checker uses a window shifting technique to improve
the level of efficiency at which it verifies proofs of unsatisfiability.
At the same time, we suggest that tree-search-based SAT solvers should use an equivalent relation encoding to emit proofs of subproblems.
In our experiments, TreeRat was able to verify  almost all proofs within 20000 seconds. On this point, TreeRat is shown to be
superior to gratgen, which is an improved version of DRAT-trim.
Also, in most cases, TreeRat is faster than gratgen. Like DRAT-trim, TreeRat can output also trace dependency graphs. Its output format is LRAT.
The correctness of TreeRat can be ensured by checking its LRAT output.


\end{abstract}

\section{Introduction}
It has been reported that SAT solvers, even some of the best, have had bugs, even during competitions \cite {DRAT,SAT:bug,SpyBug}.
Consequently, the robustness and correctness of SAT solvers are receiving more and more attention. Validating refutation proofs produced by SAT solvers
is regarded as one of the most effective approaches to verifying their robustness and correctness.
In recent years, SAT competitions have begun to check the proof of unsatisfiability emitted by SAT solvers. As far as we know, the existing proof checkers
have proven to be very slow.
The verification time often exceeds the proof generation time. For example, in the SAT Competition 2016, the verification and solving times were limited to 20000 and 5000 seconds, respectively. Even so, there were still many proofs that were not verified \cite {sc16}.
This leads to difficulty in identifying whether it is the solver or the proof checker that is buggy. Therefore, how to speed up a proof checker is a problem deserving of study.

In 2017, Cruz-Filipe et al. presented a new format, called GRIT (Generalized Resolution Trace) \cite{GRIT:17}, and then extended further it to the LRAT (Linear RAT) \cite{LRAT:17} format. In the same year, Lammich extended also GRIT. Lammich's format is called GRAT \cite{GRAT:17}, which is essentially the same as LRAT.
 Compared with the DRAT proof format, GRIT and LRAT are fast, but it is difficult for a SAT solver to output GRIT or LRAT proofs, especially for various SAT simplification procedures.
Furthermore, figuring out a resolution order needs extra work. GRIT and LRAT are essentially a variant of TraceCheck dependency graphs. In general, the TraceCheck format proof is extremely long. In the worst case, it is exponential in the size of the
conflict graph \cite{RUP,BIGsize}. This is also the fatal drawback of GRIT and LRAT.

The focus of this paper is different from that of DRAT-trim \cite {DRAT}, which focuses on extending the proof formats from DRUP to DRAT, such that proofs emitted by SAT solvers containing solving techniques such as extended resolution and blocked clause addition \cite {BCE:99}, can be verified. Rather, it is on speeding up existing proof checkers via a technique based on window shifting. This new technique can efficiently verify proofs that are emitted by tree-search-based SAT solvers but not verified by DRAT-trim. The proof checker based on the new technique is called TreeRat. It not only is fast but also can output TraceCheck dependency graphs. Our TraceCheck format is
consistent with the LRAT format. For detailed LRAT formats, see \cite{LRAT:17}.

\section{Preliminaries}

This section defines the notations and notions used throughout the paper.

A CNF (Conjunctive Normal Form) formula is defined as a finite conjunction of clauses, and also can be denoted by a finite set of clauses.
 A clause is a disjunction of literals, also written as a set of literals,
 each literal being either a Boolean variable or its complement. The
complement of a variable $x$ is denoted by $\bar{x}$ or $\neg x$.
Usually, a CNF formula $F$ is written as either $F = C_1 \wedge C_2 \wedge\cdots \wedge C_n $ or $F = \{C_1,C_2, \ldots, C_n\}$,
where $C_i (1 \leq i \leq n)$ is a clause. A clause is written as either $C = x_1 \vee x_2 \vee
\cdots \vee x_m$, or $C = \{ x_1, x_2,
\ldots, x_m\}$, where $x_i (1 \leq i \leq m)$ is a literal. The negation of $C$ is interpreted as $\overline{C} = \overline{x_1} \wedge \overline{x_2} \wedge
\cdots \wedge \overline{x_m}$. The cardinality of a set $X$ is denoted by $|X|$.
 A clause with only one literal is called a unit clause or unit literal.

Boolean Constraint Propagation (BCP) (or unit propagation) is a core component of a CDCL solver, and used also in our proof checker.
Its goal is to search for all unit clauses and simplify clauses in $F$ until there is no new unit clause.
We can achieve this goal by repeating the following process until
fixpoint: If there is a unit clause $x \in F$, remove the literal $\overline{x}$ from all clauses in $F$. Notice,
a clause $C$ will become a new unit clause if there is only one literal $y \in C$ such that for each $x \in C$ with $x \neq y$, $\overline{x}$ is a unit clause. Here is the pseudo-code of BCP.

\clearpage

\begin{small}
\begin{flushleft}
\emph{BCP} (CNF formula $F$)\\
\hskip 4mm   {\bf for } each unit clause $x \in F $ {\bf do} \\
\hskip 12mm     {\bf for } $ C \in F $ with $\bar{x} \in C$ {\bf do} \\
\hskip 20mm          $ C \leftarrow C \setminus \{\bar{x}\}$ \\
\hskip 4mm   {\bf return} $F$\\
\end{flushleft}
\end{small}

A clause $C$ is said to be a conflict clause in $\emph{BCP} (F)$ if and only if
$\overline{C} \subset \emph{BCP} (F)$. A clause $C$ with $|C|>1$ (superficially it seems not to be a unit clause) is said to be a unit
clause in $\emph{BCP} (F)$ if and only if $\overline{C} \not\subset \emph{BCP} (F)$ \& $\exists_x (\overline{C} \setminus \{x\}) \subset \emph{BCP} (F)$.
In the real implementation of BCP,
upon reaching a conflict, the process stops to search for the remaining unit clauses and then returns immediately.

There are a few approaches to proving refutations produced by SAT solvers, such as resolution proofs \cite{resolProof}, clausal proofs \cite{clausProof} etc.
This paper considers only clausal proofs. So far, clausal proofs have had two basic proof formats: RUP \cite{RUP} and RAT \cite{DRAT}, which are short for \emph{Reverse Unit Propagation} and \emph{Resolution Asymmetric Tautology}, respectively.
Clausal proofs are performed via  a sequence of \emph{inferences} ($I_1,I_2,\ldots, I_m$), which can be regarded as learned clauses produced by a SAT solver.
In RUP formats, a clause $C$ is said to be a \emph{inference} w.r.t. $F$ if \emph{BCP}($F \cup \overline{C}$) results in a conflict, i.e.,
the empty clause $\emptyset \in$ \emph{BCP}($F \cup \overline{C}$).
Inferences satisfying the above definition are also called RUP inferences.
In RAT formats, a clause $C$ is a \emph{inference} w.r.t. $F$ if and only if
either (\emph{i}) $C$ is a RUP inference or (\emph{ii}) there is a literal $l \in  C$  such that for all $D \in F$ with $\bar{l} \in D$, it holds that $\emptyset \in$ \emph{BCP}($F \cup \overline{E}$), where $E = D \setminus \{\bar{l}\} \cup C$. A proof clause with property (\emph{ii}) is also called a RAT \emph{inference}. From this definition, RAT formats are compatible with RUP formats.

In subsequent sections, we will use the term clauses to refer to clauses in an input formula, while inferences will refer to clauses in a proof,
unless otherwise mentioned.

\section{The Relation of an input Formula and its proof}

Given a CNF formula $F$ and its inference sequence ($I_1,I_2,\ldots, I_m$), the task of our proof checker is to check whether each $I_k$ ($k=1,2,\ldots,m$) is an inference w.r.t. $F \cup \{ I_1,I_2,\ldots, I_{k-1}\}$. If each check is true and $I_m = \emptyset$, $F$ is proven to be unsatisfiable. To achieve this, so far the existing proof checkers have depended heavily on the order of inferences produced by a SAT solver. They check each inference in either the forward chronological order or its reverse (backward). For example, the main mode of the checker DRAT-trim \cite{DRAT} is backward. Nevertheless, the following lemma tells us that it is not necessary to verify each inference in the  chronological order.

\begin{lemma}
For any CNF formula $F$ and any inference sequence $(I_1,I_2,\ldots, I_m)$,  $F \wedge I_1 \wedge I_2 \wedge \cdots \wedge I_m $
is logically equivalent to $F \wedge I_{\sigma(1)} \wedge I_{\sigma(2)} \wedge \cdots \wedge I_{\sigma(m)} $, where $\sigma$ is a
permutation of \{$1,2,\ldots,m$\}.
\end{lemma}

\noindent Here two formulas are logically equivalent means that they have the
same set of solutions. A permutation of a set $S$ is defined as the bijections from $S$ to itself. The above lemma indicates that
a proof checker can verify a proof in any order. It is certainly difficult to find the best order for checking. However, based on the following theorem, we can
construct a proof of a CNF formula by splitting inferences into some subsets.

\begin{theorem}
\label{theo:local}
Given a CNF input formula $F$, an inference set $S$, and a subset $T=\{J_1,J_2,\ldots, J_n\} \subset S$,  if
$\emptyset \in \emph{BCP}(F \cup \{J_1,J_2,\ldots, J_{k-1}\} \cup \overline{J_k})$ for $k=1,2,\ldots,n$, verifying proof $S$ of $F$
is equivalent to verifying proof $S \setminus T$ of $F \cup T$.
\end{theorem}
\begin{proof}
When $\emptyset \in \emph{BCP}(F \cup \{J_1,J_2,\ldots, J_{k-1}\} \cup \overline{J_k})$ for $k=1,2,\ldots,n$,
it is easy to prove that $F$ is logically equivalent to $F \cup T$. But not only that, $F \cup T$ can be regarded as an input formula.
By Lemma 1, $F \cup T \cup (S \setminus T)$ is logically equivalent to $F \cup S$. Hence, this theorem has been justified.
\end {proof}

We will apply this theorem to extract unit clauses from given inferences and add the extracted unit clauses to the
input formula to speed up the subsequent verification. See procedure \emph{UnitProbe} in the next section. We can also use  this theorem to extract general
clauses including binary or ternary clauses from inferences.

  Although inferences produced by a solver can present conflict clauses in BCP, the following theorem shows that all conflict clauses should be in the input formula,
  not in the inferences.

\begin{theorem}
Given a CNF formula $F$, and an inference set $\{I_1,I_2,\ldots, I_m$\}, let $P_i=F \cup \{I_1,I_2,\ldots, I_{i-1}\}$. Suppose $1 \leq j < k \leq m$, if $I \in P_j$ is a conflict clause in $\emph{BCP}(P_k \cup \overline{I_k})$, and $\emptyset \in \emph{BCP}(P_i \cup \overline{I_i})$ holds for $i=1,2,\ldots,k$, then there exists a clause $C \in F$, such that $C$ is a conflict clause in $\emph{BCP}(P_k \cup \overline{I_k})$.
\end{theorem}

\begin{proof}
We will prove this by induction on $k$. When $k=1$, clearly, a conflict clause in $\emph{BCP}(F \cup \overline{I_1})$ is in $F$.
Suppose it is true for $k<n$ that when $k=n$, by theorem hypothesis, there exists $j < n$ such that $I_j$ is a conflict clause in $\emph{BCP}(P_n \cup \overline{I_n})$, i.e., $\overline{I_j} \subset \emph{BCP}(P_n \cup \overline{I_n})$. Using the fact that for $i=1,2,\ldots,k$, $\emptyset \in \emph{BCP}(P_i \cup \overline{I_i})$, we have $\emptyset \in \emph{BCP}(P_j \cup \overline{I_j})$. By induction hypothesis, we have that there exists $C \in F$ such that $\overline{C} \subset \emph{BCP}(P_j \cup \overline{I_j})$. It follows that $\overline{C} \subset \emph{BCP}(P_n \cup \overline{I_n})$.
By the principle of induction, it is true for all $k$.
\end {proof}

\section {A Proof Checker Based on Window Shifting}

Our proof checker, TreeRat, supports RAT formats compatible with RUP formats and verifies a proof of unsatisfiability in the reverse of chronological order.
It requires two input files: a formula and a proof. Each clause in a proof is called an inference. We prepare two mark variables for each inference $I$,
which are denoted by $I.verfied$ and $I.used$. Initially, all the mark variables are set to {\bf false}. When inference $I$ is used as a unit or conflict
clause in \emph{BCP}, $I.used$ is set to {\bf true}. Once $I.used$ becomes {\bf true}, inference $I$ must be verified. Otherwise, we skip it to save checking
time. We mark the verified inference $I$ by setting $I.verfied$ to {\bf true}.

  Inferences can be classified as RUP or RAT. The proof checker TreeRat focuses on how to speed up the RUP inference check. Its module for checking RUP
  inferences consists of two subroutines: \emph{UnitProbe} and \emph{WindowShiftCheck}.

\emph{UnitProbe} corresponds to a probing failed literal procedure in CDCL solvers.  Its function is to extract independently unit clauses from inference set $S$ and add them to input formula $F$. This can be done by detecting whether a unit clause $x$ in $S$ satisfies $ \emptyset \in BCP(F \cup \bar{x})$.
According to Theorem \ref{theo:local}, as long as $\emptyset \in BCP(F \cup \bar{x})$, adding inference $x$ to $F$ is valid.
Here is the pseudo-code of \emph{UnitProbe}.

\begin{small}
\begin{flushleft}
\emph{UnitProbe} (CNF formula $F$, inference set $S$)\\
\hskip 4mm   {\bf for } each unit clause $x \in S $ {\bf do} \\
\hskip 12mm     {\bf if } $ \emptyset \in BCP(F \cup \{\bar{x}\})$ {\bf then} \\
\hskip 20mm          $ S \leftarrow S \setminus \{x\}$ \\
\hskip 20mm          $ F \leftarrow F \cup \{x\}$ \\
\hskip 4mm   {\bf return} $\langle F, S\rangle$\\
\end{flushleft}
\end{small}

Next, we introduce a verifying technique based on window shifting. This technique uses the locality of inferences to check the validity of each inference in the range of at most $\theta$ inferences, rather than the range of the whole set of inferences. We name a procedure implementing such a verification as \emph{WindowShiftCheck}. Removing parameter $\theta$, \emph{WindowShiftCheck} is exactly the same procedure as
that used by the usual RUP proof checker in the backward mode. Here is its pseudo-code.

\begin{small}
\begin{flushleft}
\emph{WindowShiftCheck} (CNF formula $F$, inference set $\{I_1,\ldots,I_m\}$, window size $\theta$)\\
\hskip 4mm   {\bf for} $i=m$ {\bf down to} $1$ {\bf do}\\
\hskip 12mm     {\bf if}  $I_i.verified$={\bf true or} $I_i.used$={\bf false then continue} with next $i$\\
\hskip 12mm     $ P_i=F \cup \{I_{i-\theta},\ldots,I_{i-1}\} \cup \overline{I_i}$ \\
\hskip 12mm     {\bf if} $ \emptyset \notin BCP(P_i)$ {\bf then continue} with next $i$\\
\hskip 12mm     {\bf for} each $I_t$ with $ 0<t$ \& $i-\theta < t <i$  {\bf do}\\
\hskip 20mm           {\bf if } $\exists_x (\overline{I_t} \setminus \{x\}) \subset \emph{BCP} (P_i)$ {\bf then} $I_t.used \leftarrow$ {\bf true} \\
\hskip 12mm     $I_i.verified \leftarrow $ {\bf true} \\
\end{flushleft}
\end{small}

Notice, the above $P_i$ is an approximate expression. Let $Q_i=\{I_t | 0<t$ \& $i-\theta < t <i\}$.  Its exact expression is $ P_i=F \cup Q_i \cup \overline{I_i}$. Let $|S|$ be the total number of inferences in a proof. When $\theta < |S|-1$,  this procedure cannot ensure that every verification is successful.
Therefore, we invoke at least one time this procedure with $\theta=|S|$. In general, the larger $|Q_i|$, the slower the speed of \emph{WindowShiftCheck}.
To speed up it, we can remove inferences of size $> \mu$ from $Q_i$. $\mu$ is usually set to 6. Using parameters $\theta$ and $\mu$, we may compute $P_i$ by the following pseudo-code.

\begin{small}
\begin{flushleft}
\hskip 12mm     $ P_i=F \cup \overline{I_i}$ \\
\hskip 12mm     {\bf for} each $t$ with $ 0<t$ \& $i-\theta < t <i$ {\bf and} $|I_t| \leq \mu$ {\bf do}\\
\hskip 20mm           {\bf if } $\theta = \infty$ {\bf or} $\exists_{x,y}(\overline{I_t} \setminus \{x,y\}) \subset \emph{BCP} (P_i)$ {\bf then} $P_i \leftarrow P_i \cup \{I_t\}$ \\
\end{flushleft}
\end{small}
\noindent In the above pseudo-code, $\mu$  is set to $\infty$ when $\theta = \infty$, 6 otherwise. Clearly, when $\theta = \infty$, the above pseudo-code results in $ P_i=F \cup \{I_1,\ldots,I_{i-1}\} \cup \overline{I_i}$, which corresponds to the slowest mode,
but lets nothing escape from the checking net.

Now we describe the basic idea of the proof check \emph{TreeRat} as follows. It first performs RUP verification via the subroutines given above, and then performs RAT verification. Let $S=\{I_1,\ldots,I_m\}$) be an inference set. In RUP verification, we first use \emph{UnitProbe} to add unit clauses in $S$ verified independently to input formula $F$.
Then we make two calls to \emph{WindowShiftCheck}. One is to verify each inference on a very small scale. The other is to verify each inference in the whole range. The first call is done in an approximate way. The window size $\theta$ is set to 10000. The second call is done in an exact way. The window size $\theta$ is set to $\infty$. Here is the pseudo-code of \emph{TreeRat}.

\begin{small}
\begin{flushleft}
\emph{TreeRat} (CNF formula $F$, inference set $S=\{I_1,\ldots,I_m\}$)\\
\hskip 4mm    $\langle F, S\rangle \leftarrow $ \emph{UnitProbe}($F$, $S$)\\
\hskip 4mm     {\bf if} $\emptyset \notin$ \emph{BCP}($F \cup S)$ {\bf return failure} \\
\hskip 4mm     $ S' \leftarrow \{ I_t | I_t \in S$ \& $(t>100000$ {\bf or} $|I_t| \leq 6)$\} \\
\hskip 4mm     \emph{WindowShiftCheck}($F$, $S'$, 10000)\\
\hskip 4mm     {\bf if} $\exists_I I.verfied$={\bf false} \& $I.used$={\bf true then}\\
\hskip 12mm            \emph{WindowShiftCheck}($F$, $S$, $\infty$)\\
\hskip 12mm            {\bf if} $\exists_I I.verfied$={\bf false} \& $I.used$={\bf true then} \emph{RATcheck}($F$,$S$)\\
\end{flushleft}
\end{small}

In the above pseudo-code, procedure \emph{RATcheck} is used to check whether the unverified inferences are a RAT inference.
Here we omit the description of \emph{RATcheck}, because it is the same as the RAT check of \emph{DRAT-trim}. For more information on its
implementation, see Section 7 in \cite {RATextend}. Compared to a RUP check, a RAT checker is less efficient. This is because
 a RAT check needs to maintain a full occurrence list of all clauses containing the
negation of the resolution literal, while a RUP check can use a two watch pointer data structure.
 Notice, we do not run a RAT check until the end of all the RUP inference checks. In general, most inferences can be validated using the RUP check.
 The remaining inferences should be small. If they are small, building a literal-to-clause lookup table is not so expensive in order to run a RAT check.
 One of differences between our checker and \emph{DRAT-trim} is that \emph{DRAT-trim} combines a RUP check and a RAT check,
 whereas we separate them.  To build a literal-to-clause lookup table, \emph{DRAT-trim} scans
 the current formula many times, while we scan it only once.

  During the verification, for any inference $I_j$, if there exists an inference $I_k$ ($j < k$) that is a unit literal $x$ such that $x \in I_j$,
  we let $I_j$ become inactive, i.e., remove $I_j$ from the two-watch-pointer data structure. This strategy speeds up verification.  Of course,
  when verifying a DRAT format proof, in most cases, such an inference can be removed from the watch list by a deletion step.

  To find efficiently deleted clauses (inferences),  proof checkers need a hash function.  TreeRat uses a hash function
  in a manner different from that of DRAT-trim.  The hash used in TreeRat is a weighted sum. In detail, given a clause $C$ with $m$ literals, and supposing $C =\{x_1, x_2, \ldots, x_m\}$,
  we sort literals in $C$ to find a permutation $\sigma$ of \{$1,2,\ldots,m$\} satisfying
    $x_{\sigma(1)} \leq x_{\sigma(2)} \leq \cdots \leq x_{\sigma(m)}$. Using this permutation $\sigma$, TreeRat defines a hash function as hash$(C) = m+\sum_{i=1}^{m} x_{\sigma(1)} \times i$.

  We noted that too many binary and ternary inferences slow down the verification process. Most of them will not be used. For this reason, we remove some of the
  binary and ternary inferences from the watch list. Once some inference verification fails, we restore them partially.

   During SAT Competition 2014, a few runs reportedly generated proofs of over 100 GB \cite {sc16}. It is difficult to store all the data of such huge proofs in
   the main memory of the general PC platform. For this reason, we store only active inferences, not all inferences, in the memory.
   When an inference switches from inactive status to active status, we load it from the proof file.

\section{Empirical evaluation}

This experiment verifies the proofs outputted  by two SAT solvers: abcdSAT rat and Glucose 4.0 \cite{Glucose}.
 AbcdSAT rat is the improved version of abcdSAT drup entering the SAT Competition 2016.
The search engine of the two versions is completely the same. The difference between them is that they use different proof formats to generate a proof of unsatisfiability. On subproblems obtained at tree search depth $n > 1$, abcdSAT rat produces inferences with DRAT formats,
while abcdSAT drup uses DRUP formats to produce inferences. In some cases,
abcdSAT divides the original problem into multiple subproblems, using a tree-based search mechanism. In such a search mechanism, given an original problem $F$ and
branch literals $x_1,x_2, \ldots, x_n$, a subproblem with depth $n$ is defined as \emph{BCP}($F \cup \{x_1 \wedge x_2 \wedge \cdots \wedge x_n \}$), i.e., a formula
resulting from unit propagations $x_1,x_2, \ldots, x_n$ on $F$.
     Suppose $I= y_1 \vee {y_2} \vee \cdots \vee y_m$ is an inference on subproblem \emph{BCP}($F \cup \{x_1 \wedge \cdots \wedge x_n \}$).
$I$ corresponds to an inference $y_1 \vee \cdots \vee y_m \vee \overline{x_1} \vee \cdots \vee \overline{x_n}$ on the original problem $F$.
According to this correspondence  rule, abcdSAT drup must transfer inferences on a subproblem into inferences on the original problem.
Thus, the proof emitted by abcdSAT drup contains a large amount of redundant information so that the speed of the checker is very slow.
To reduce the redundant information, as an output of a proof, abcdSAT rat transfers inference $I$ to $\bar {z} \vee y_1 \vee \cdots \vee y_m $, where $z$ is an auxiliary
variable, which is defined as $z=x_1 \wedge x_2 \wedge \cdots \wedge x_n$. When generating subproblem \emph{BCP}($F \cup \{x_1 \wedge \cdots \wedge x_n \}$),
as a part of a proof, abcdSAT rat must produce the following $n+1$ RAT inferences:

\[\left \{ \begin {array}
         {l@{\quad \quad}l}
         z \vee \overline{x_1} \vee  \overline{x_2} \vee \cdots \vee \overline{x_n} \\
         {\bar z} \vee  x_i & 1 \leq i \leq n
         \end {array} \right . \]
\noindent Using this encoding output technique, we found that the total number of literals of inferences decreases sharply for formulas suitable to the tree-based search.

In general, in order to verify RAT inferences, a RAT checker needs to maintain a full occurrence list of all clauses. Nevertheless, we noted that verifying the
RAT inferences that denote the equivalent relation
need not mean maintaining
a full occurrence list. We can just follow $ z \vee \overline{x_1} \vee  \overline{x_2} \vee \cdots \vee \overline{x_n}$ to produce all clauses ${\bar z} \vee  x_i$. When inputting such inferences, our proof checker can verify them by checking whether the negation of each $x_i$ in ${\bar z} \vee  x_i$ is contained in the first clause with the pivot $z$. In other words, we need not run the additional RAT check. Therefore, the checking cost
 of RAT inferences denoting equivalent relation is very cheap.

\begin{table}
\caption{Proof timing comparisons on special instances. Time is CPU time in seconds.} \label{timeTab}
\begin{center}
\begin{tabular}{|l|c|c|c|c|c|c|}
\hline  \hline
\multicolumn{1}{|c|}{SAT 2016}  & abcdSAT rat  & TreeRat shift & DRAT-trim & TreeRat no shift & gratgen \\
\multicolumn{1}{|c|}{Instances} & solving time & proof time & proof time  &  proof time & proof time \\
\hline
ctl\_4291\_567\_8\_un      &  675.7  & 1725.9    &  $>20000$ & 2141.9   &  19041.7 \\
ablmulub8x16o              & 3013.8  &  $>20000$ &  $>20000$ & $>20000$ &  $>20000$ \\
hitag2-8-60-0-94           &  1996.1 &  9665.7   &  $>20000$ & 13987.6  &  14768.5 \\
eq.atree.braun.12          & 2011.5  & 10502.8   & $>20000$  & $>20000$ &  $>20000$ \\
Schur\_161\_5\_d40         & 1252.7  & 4126.3    &  $>20000$ & 7046.6   &  7160.8 \\
Schur\_161\_5\_d42         & 386.9   & 981.6     &  3065.9   & 1654.7   &  1701.7 \\
Schur\_161\_5\_d43         & 219.8   & 612.3     &  2033.5   & 1241.2   &   1294.1 \\
\hline
\end{tabular}
\end{center}
\end{table}

\begin{figure}
\centering
\includegraphics[height=7.2cm]{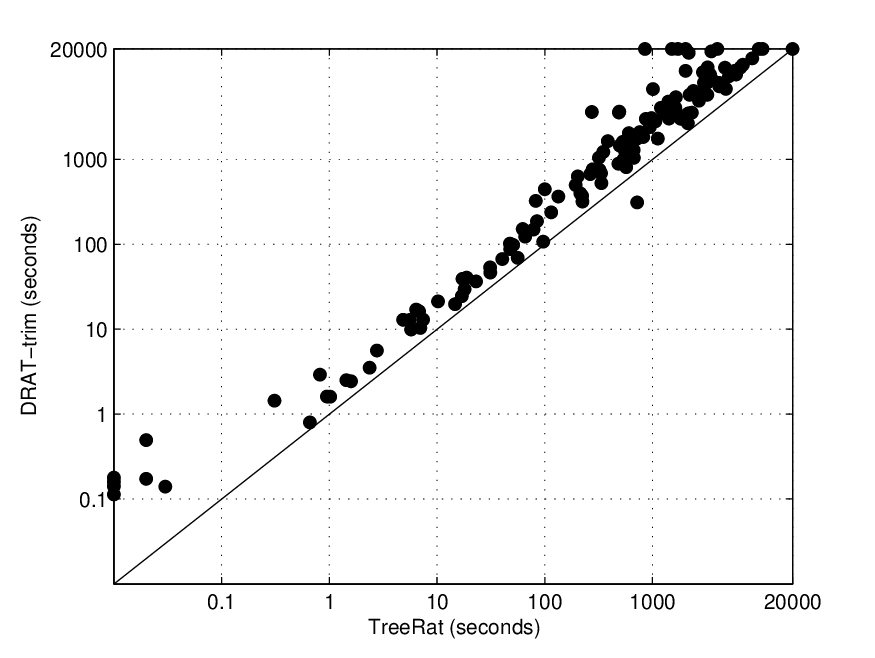}
\caption{Comparing the runtimes of \emph{DRAT-trim} and \emph{TreeRat shift}
on 131 proofs emitted by abcdSAT.} \label{scaterFig}
\end{figure}

\begin{figure}
\centering
\includegraphics[height=7.2cm]{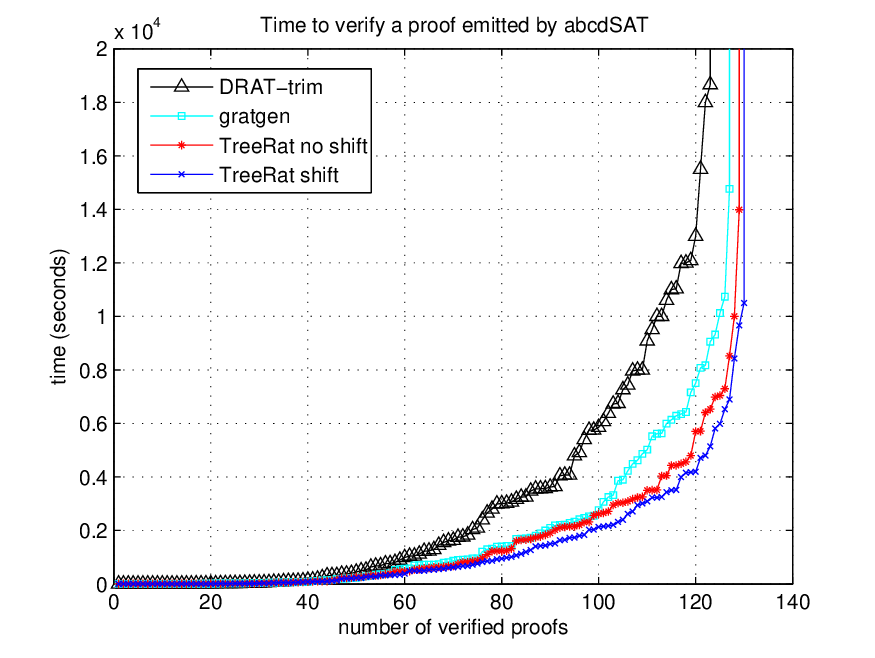}
\caption{The number of proofs that \emph{DRAT-trim}, \emph{gratgen} and two versions of \emph{TreeRat}
are able to verify in a given amount of time. The proofs were emitted by abcdSAT. The $x$-axis
denotes the number of verified instances, while the $y$-axis denotes
running time in seconds. } \label{cactusFig}
\end{figure}

\begin{figure}
\centering
\includegraphics[height=7.2cm]{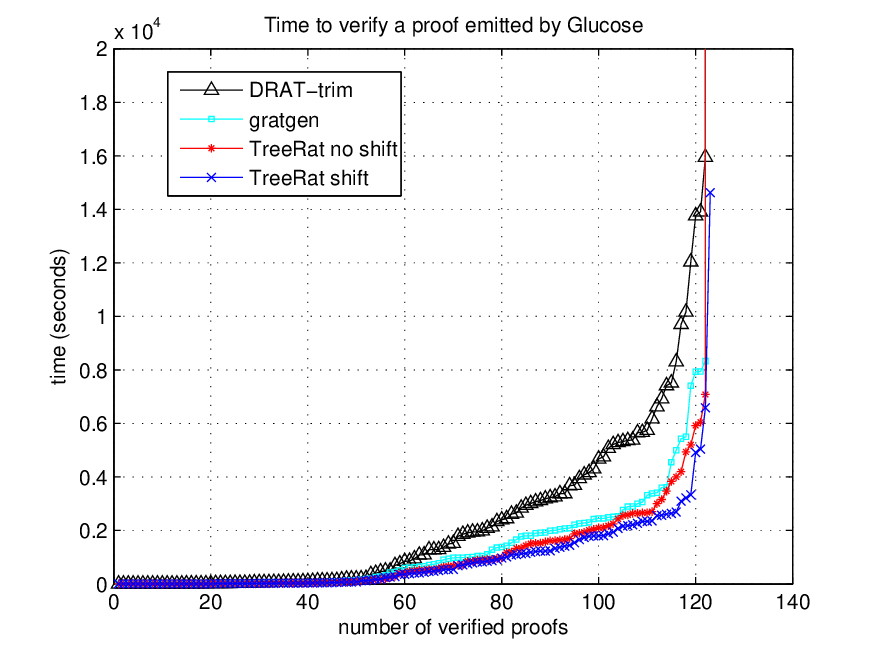}
\caption{The number of proofs that \emph{DRAT-trim}, \emph{gratgen} and two versions of \emph{TreeRat}
are able to verify in a given amount of time. The proofs were emitted by Glucose.} \label{glucoseFig}
\end{figure}


SAT solvers and proof checkers in the experiment were run under the following
platform: Intel core i5-4590 CPU with speed of
3.3 GHz. The timeout for each proof checker was set to 20000 seconds.
 Each software is written in either C or C++.
   To be able to check the correctness of verification given by \emph{TreeRat}, we added a tool that outputs trace dependency graphs in LRAT formats \cite {LRAT:17}.   LRAT formats used here support RAT proofs.  In addition,
we developed the verification tool \emph{tracecheck\_LRAT} to check the trace file outputted by \emph{TreeRat}.
Thus, the correctness of \emph{TreeRat} can be ensured by these tools.
 The source codes of \emph{abcdSAT rat}, \emph{TreeRat} and \emph{tracecheck\_LRAT} are available at
https://github.com/jingchaochen.

Lammich's \emph{gratgen} is a tool to verify DRAT unsatisfiability certificates and convert them into GRAT certificates \cite {GRAT:17}.
\emph{DRAT-trim} and \emph{gratgen} both use a core-first unit propagation policy. The main difference between them is that \emph{DRAT-trim} uses a single watchlist,
 while \emph{gratgen} does two separate watchlists, one for the core and one for the non-core inferences. Whenever a inference is used in a conflict analysis, it is moved from the non-core to the core watchlists. \emph{TreeRat} adopts also the core-first unit propagation policy, but uses three separate watchlists, one for the core, one for the binary non-core and one for the non-binary inferences. In addition, we clear periodically the core watchlist. In general, whenever the size of the core watchlist exceeds 70000, we move all the core to the non-core watchlists. This can be considered as a part of window shift operations. \emph{TreeRat} is not only faster
 than \emph{DRAT-trim} and \emph{gratgen}, but also uses less memory. In our experiment, the memory requirement of \emph{TreeRat} does not exceed 4 GB, while \emph{DRAT-trim} and \emph{gratgen} both exceeds 20 GB.

Table~\ref{timeTab} shows the performance of four checkers on a few proofs emitted by \emph{abcdSAT rat}.
The original formulas that are used to produce these proofs are from the SAT competition 2016. Notice, these proofs were not verified by \emph{DRAT-trim} at that time \cite {sc16}. To examine the effect of the window shifting technique, we used two versions of \emph{TreeRat} with and without window shifting. \emph{TreeRat shift} denotes
TreeRat with window shifting. In our experiment, the window shifting size was generally set to 1000. TreeRat without window shifting is denoted by \emph{TreeRat no shift}. As shown in Table~\ref{timeTab},  although both of \emph{TreeRat shift} and \emph{no shift} are faster than \emph{DRAT-trim},
 the \emph{shift} version is faster. Moreover, the number of proofs verified by \emph{TreeRat shift} is one more than that of proofs verified by \emph{TreeRat no shift}.
The reason the former is faster than the latter on some proofs is that when verifying each subproblem proof, \emph{TreeRat shift} uses the window shifting technique to remove the inactive subproblem proofs from watchlists. This is also the advantage of the window shifting technique. The last column shows the time required by \emph{gratgen}. Based on our experimental observation, \emph{gratgen} is slower than \emph{TreeRat shif}, and close to \emph{TreeRat no shift}.

In addition to the 7 proofs shown in Table~\ref{timeTab}, we tested the verification of the other proofs emitted by
\emph{abcdSAT rat} and that of proofs emitted by \emph{Glucose}. In total, we generated 131 proofs and 123 proofs by \emph{abcdSAT rat} and \emph{Glucose}, respectively.
All the instances in the experiment are from the SAT Competition 2016.


Figure~\ref{scaterFig} shows a log-log scatter plot comparing the running times
of \emph{TreeRat shift} and \emph{DRAT-trim} on the 131 proofs. Each point corresponds to a given proof. A
point at line $y=20000$ (resp., $x=20000$) means that the proofs on that point were not
verified by \emph{DRAT-trim} (resp., \emph{TreeRat shift}). As shown in Figure~\ref{scaterFig}, almost all the points
appear over the diagonal. This means that, in almost all the cases, \emph{TreeRat shift} is faster than
\emph{DRAT-trim}.

  Figures \ref{cactusFig} and \ref{glucoseFig} show a cactus plot related to the performance comparison of the four proof checkers on proofs emitted by
  \emph{abcdSAT rat} and \emph{Glucose}, respectively. The two cactus plots show clearly that two versions of \emph{TreeRat} outperform \emph{DRAT-trim} and \emph{gratgen}.  \emph{DRAT-trim} is the slowest.
  As shown in the two figures, the \emph{TreeRat shift} and \emph{TreeRat no shift} curve are almost always below the\emph{gratgen} curve. That
is, in a given amount of time, two versions of \emph{TreeRat} almost always verified more proofs than \emph{gratgen} did.
The reason why \emph{TreeRat no shift} is also faster than \emph{gratgen} is that the watchlist of \emph{TreeRat no shift} is different from that of \emph{gratgen}.
 It is easy to see that the improvement shown in
Figure~\ref{glucoseFig} is smaller than that shown in Figure~\ref{cactusFig}. This is because abcdSAT is a SAT solver based on a tree search, while Glucose has no
tree search mechanism.  TreeRat is better at verifying proofs emitted by a tree-search-based SAT solver.

\section {Conclusions}

In this paper, we have described a proof checker called TreeRat, which we developed and which is more efficient than previous checkers.
  However, compared to the solving efficiency of SAT solvers, TreeRat is still inefficient. Consequently,
  the following problems arise: (\emph{i}) What is the most efficient proof checker? (\emph{ii}) To prove refutations produced by SAT solvers,
   does there exist an approach better than clausal proofs? Clausal proofs are easily  produced, but require much more disk space, and their verification is time consuming.
   What is the trade off between proof production difficulty and verification efficiency? How to resolve the trade off is well worth studying.

\section*{Acknowledgements}

The author is grateful to students at writing center of Stanford University who helped to improve
the language of the paper, and to Nigel Horspool who gave helpful discussions and tested the experiments.

\bibliographystyle{splncs}
\bibliography{treeRat}

\end {document}